\documentclass[twocolumn,preprintnumbers,amsmath,amssymb]{revtex4}
\input psfig.sty  
\def \be {\begin{equation}} 

\def \ee {\end{equation}} 
\def \bea {\begin{eqnarray}} 
\def \eea {\end{eqnarray}} 

\usepackage{graphicx}
\usepackage{dcolumn}
\usepackage{bm}
\usepackage{epsfig}

\newcommand*{\ltsim}{\ {\raise-.75ex\hbox{$\buildrel<\over\sim$}}\ }
\newcommand*{\gtsim}{\ {\raise-.75ex\hbox{$\buildrel>\over\sim$}}\ }
\newcommand*{\proptosim}{\ {\raise-.75ex\hbox{$\buildrel\propto\over\sim$}}\ }

\begin{document}

\title{A possible analogy between the dynamics of a skydiver and a scalar field:\\ cosmological consequences}

\author{F. E. M. Costa} \email{ernandesmatos@ufersa.edu.br}
\affiliation{Universidade Federal Rural do Semi-\'Arido, 59900-000, Pau dos Ferros, RN, Brazil}

\date{\today}

\begin{abstract}

The cosmological consequences of a slow rolling scalar field with constant kinetic term in analogy to the vertical movement of a skydiver after reaching terminal velocity are investigated. It is shown that the terminal scalar field hypothesis is quite realistic. In this approach, the scalar field potential is given by a quadratic function of the field. This model provides solutions in which the Universe was dominated in the past by a mixture of baryons and dark matter, is currently accelerating (as indicated by type Ia supernovae data), but will be followed by a contraction phase. The theoretical predictions of this model are consistent with current observations, therefore, a terminal scalar field is a viable candidate to dark energy. 

\end{abstract}

\maketitle

\section{Introduction}

The analysis and interpretation of several sets of observational data such as cosmic microwave background (Spergel et al. 2007), large scale structure surveys (Eisenstein et al. 2005) and type Ia supernovae (Permulter et al. 1998; Riess et al. 1998), when combined, indicate that the current observed Universe has a spatially flat geometry and its expansion is accelerating.
By assuming that the matter fields (baryonic matter, dark matter and radiation) account $\simeq 0.3$, in units of the critical density, of the cosmic composition and which gravity is described by general relativity theory; these results can be explained if we admit the existence of an exotic component, the so-called dark energy. The origin and nature of this component constitutes a complete mystery and represents one of the major challenges not only to cosmology but also to our understanding of fundamental physics (Sahni \& Starobinsky, 2000). 

The simplest possibility to describe the dark energy is the vacuum energy density of all existing fields in the Universe, that acts in Einstein field equations as cosmological constant $(\Lambda)$, and is equivalent to an isotropic and homogeneous fluid with a constant equation of state $\omega \equiv p_{\Lambda}/\rho_{\Lambda} = -1$. Although this possibility is in good agreement with almost all sets of observational data, the value of ${\Lambda}$ inferred by observations ($\rho_{\Lambda} = \Lambda/8\pi G \sim 10^{-47} {\rm{GeV}}^4$) is between 50-120 orders of magnitude below estimates given by quantum field theory ($\rho_{\Lambda} \sim 10^{71}$ ${\rm{GeV}}^4$). This large discrepancy originates an extreme fine-tuning problem (cosmological constant problem), and requires a complete
cancellation from an unknown physical mechanism (Weinberg 1989; Costa \& Alcaniz 2010; Costa, Alcaniz \& Deepak 2012). Furthermore, the cosmological term has a coincidence problem, which is to understand why $\rho_{\Lambda}$ is not only small, but also of the same order of magnitude of the energy density of cold dark matter today (Amendola 2000; Chimento et al. 2003; Jesus et al. 2008; Costa 2010).  

Still in the context of relativistic cosmology,
there are another possibilities to describe the dark energy component and one of the most pleasing to the theoretical is a scalar field characterized by a Lagrangian density of the type $ {\cal{L}} = (1/2) \partial_{\mu}\phi \partial^{\mu}\phi - V(\phi)$, that involve an arbitrary potential $V(\phi)$ and that is basically which determines the current dynamics of a given model. Several forms for the potential have been suggested in the literature (Ratra \& Peebles 1988; Wetterich 1995; Dodelson, Kaplinghat \& Stewart 2000) and although some of them have some attractive properties there is no reason to choose a certain potential. The main requirement that any potential must satisfy is at early times the energy density of the field is much less than the energy density of radiation (Weinberg 2008).

Scalar field models have received much attention over the past few years because it is believed that fundamental physics provides motivation for light scalar fields in nature. Thus, a quintessence field may not only be identified as the dark component dominating the current cosmic evolution, but also as a bridge between an underlying theory and the observable structure of the Universe. In this way, a considerable effort has been carried in understanding the role of quintessence fields on the dynamics of the Universe (Caldwell, Dave \& Steinhardt 1998; Carroll 1998; Carvalho et al. 2006).

In this paper, instead of proposing a new potential and investigate its cosmological consequences, we use an analogy between the vertical movement of a skydiver after reaching terminal velocity and a rolling quintessence field with constant kinetic term to derive the scalar field potential. We show that the terminal scalar field hypothesis is quite realistic. We also find cosmological solutions in which the Universe was dominated in the past by a mixture of baryons and dark matter, is currently accelerating (as indicated by type Ia supernovae data), but will be followed by a contraction phase. Finally, we apply two different cosmological tests, the distance modulus of type Ia supernovae and the BAO/CMB ratio, to compare the predictions of our model with current observational data.

\section{The model}
\vspace{-0.3cm}
The action that describes gravity, the scalar field, the matter and radiation fluids is given by

\begin{equation}\label{acao}
S = \int d^{4}x \sqrt{-g} \left[\frac{R}{2} - \frac{1}{2}(\partial \phi)^{2} - V(\phi) + {\cal{L}}_m \right]\;, 
\end{equation}
where $R$ is the Ricci scalar, $g$ is the determinant of the metric $g^{\mu \nu}$, ${\cal{L}}_m$ is the Lagrangian density of all matter and radiation fields and we have assumed $8\pi G = 1$.

For a flat, homogeneous and isotropic universe the previous action provides the following equations of motion:
\begin{equation}\label{klein}
\ddot{\phi} +3H\dot{\phi} + \frac{dV}{d \phi} = 0\;,
\end{equation}
\begin{equation}
\dot{\rho}_f + 3H(\rho_f + p_f) =0\;,
\end{equation}
\begin{equation}\label{fri}
{{H}}= \frac{1}{\sqrt{3}} \left[\frac{1}{2}\dot{\phi}^{2} + V(\phi) + \rho_f \right]^{1/2}\;,
\end{equation}
where a dot denotes a derivative with respect to $t$, $\rho_f$ and $p_f$ are, respectively, the energy density and pressure of the cosmic fluid (radiation, baryons and dark matter) and $H$ is the Hubble parameter.

\subsection{Motivation}

From a dynamic perspective, Eq. (\ref{klein}) is similar to the equation of a particle of unit mass with one-dimensional coordinate $\phi$, moving in a potential $V(\phi)$ with a frictional force $- 3H \phi$ (Weinberg 2008). In this case the expansion of the Universe acts like a frictional term on the rolling scalar field down potencial. 

Scalar fields have been considered to solve the problems of the big bang model, since there is a primordial scalar field in the Universe, it is reasonable to think that after a long time the expansion leads it to terminal regime, i.e., $\dot{\phi} \longrightarrow constant$.

\begin{figure}[t]
\centerline{\psfig{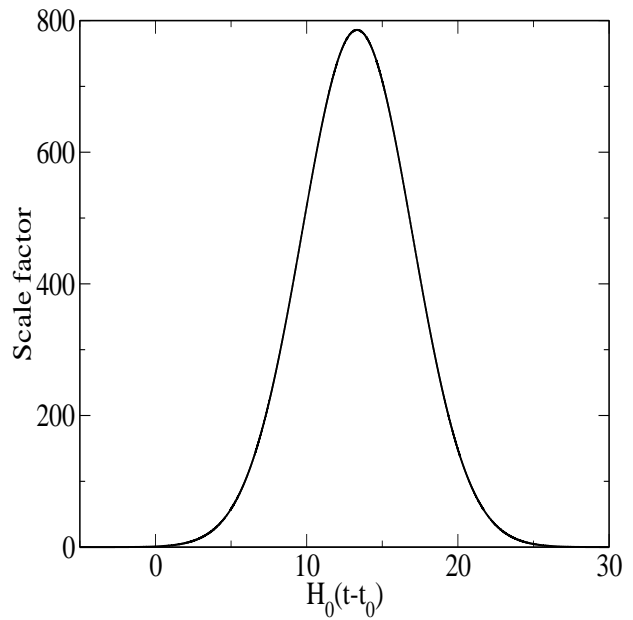}}
\caption{Evolution of the cosmic scale factor as a function of $H_0 (t- t_0)$ for $\Omega_c = 0.05$. Note that the function has a peak in which the cosmic scale factor begins to decrease and the Universe collapses.}
\end{figure}

\subsection{A terminal scalar field dominated universe}

Firstly, let us consider a universe fully dominated by a slow rolling scalar field with constant kinetic term, 
\begin{equation}
\dot{\phi} = constant \equiv  \dot{\phi}_{c} \;.
\end{equation}
In this approach, Eqs. (\ref{klein}) and (\ref{fri}) can be combined to give
\begin{equation}\label{int}
\sqrt{3}\dot{\phi}_c \left[\frac{1}{2}\dot{\phi}_{c}^{2} + V\right]^{1/2} + \frac{dV}{d\phi} = 0\;.
\end{equation}
Performing the integration of above equation, we derive the following potential
\begin{equation}\label{potential}
V(\phi) = V_0 - \sqrt{0.5\dot{\phi}_{c}^{2} + V_0}\lambda (\phi - \phi_{0}) + \frac{1}{4}\lambda^{2} (\phi - \phi_{0})^{2}\;,
\end{equation}
where $\lambda \equiv \sqrt{3}\dot{\phi}_{c}$ and $V_0$ the value of the potential to $\phi = \phi_{0}$. Potential with quadratic dependence on field has been used to model chaotic inflation scenarios, since that it presents a residual term is also possible to explain the current cosmic acceleration (Liddle, Pahud \& Urena-Lopez 2008).

\begin{figure*}
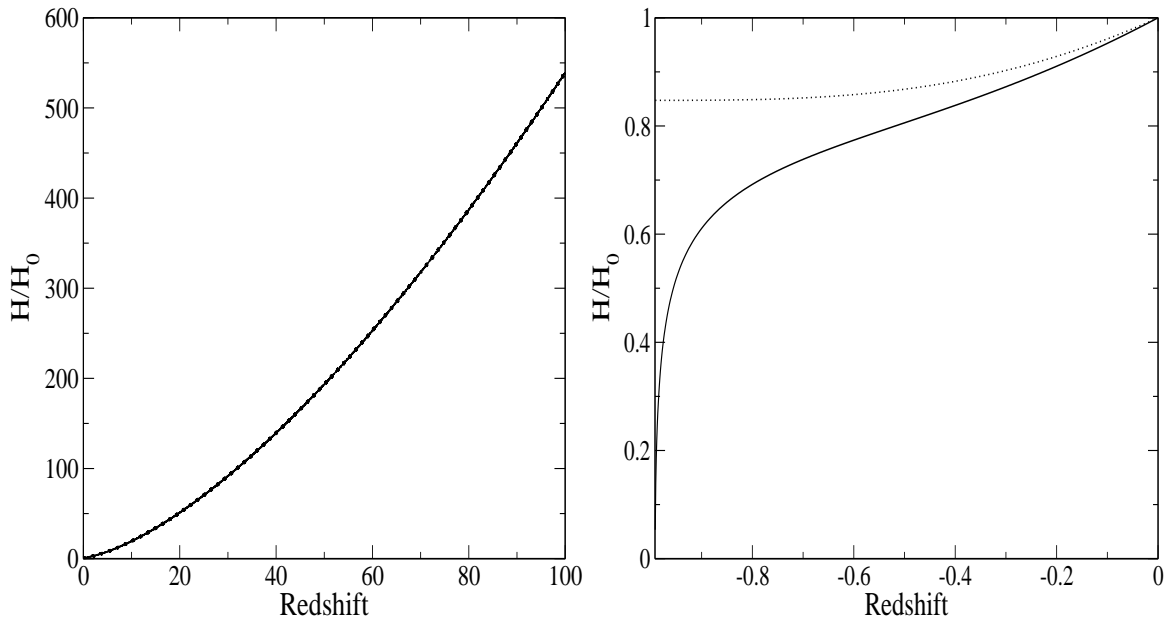

\centerline{\psfig{figure=fig2.eps,width=3.0truein,height=3.2truein}
\psfig{figure=fig3.eps,width=3.0truein,height=3.2truein}}
\caption{Evolution of the normalized Hubble parameter as a function of redshift $(z)$ for $\Omega_m = 0.282$ and $\Omega_c = 0.05$ (from past to present $-$ left panel) and (from present to future $-$ right panel). In both cases the solid line represents the model given by Eq. (\ref{taxat}) while dotted line is the standard model. Note that the evolution of the Universe from past to present is the same for both models, however, the future evolution of the two scenarios are very different.}
\label{fig:qzw}
\end{figure*}

Now, remembering that
\begin{equation}
\dot{\phi}_{c} = \frac{d\phi}{dt} \Longrightarrow \phi - \phi_{0} = \dot{\phi}_{c}(t - t_0)\;.
\end{equation}
By combining this result with Eqs. (\ref{fri}) and (\ref{potential}), one finds
\begin{equation}
a = a_0 exp[k_1 (t -t_0) - k_2 (t -t_0)^{2}]\;,
\end{equation}
where $k_1 \equiv \sqrt{(0.5\dot{\phi}_{c}^{2} + V_0 )/3} = H_0 $ and $k_2 \equiv \dot{\phi}_{c}^{2}/4$. Deriving above equation with respect to time and using the definition of $H \equiv \dot{a}/a$, one can show that
\begin{equation}\label{ht}
H = H_0 - 2k_2(t - t_0)\;.
\end{equation}
For,
\begin{equation}
t = t_0 + \frac{2}{3H_0 \Omega_{c}} \Longrightarrow H = 0\;,
\end{equation}
where $\Omega_{c} \equiv \dot{\phi}_{c}^2/3H_{0}^{2}$, the Universe will begin to collapse (see figure 1). In this case, the cosmic contraction acts like an anti friction term providing energy for the field move up potential. Note that the collapse time depends only on the kinetic term $\dot{\phi}_{c}$ and $H_0$. For $\Omega_{c} = 0.05$ and $H_0 = 68$ kms$^{-1}$Mpc$^{-1}$, one otains $t - t_0 = 190$ Gyears.

\subsection{Terminal scalar field hypothesis}

Let us now show that terminal scalar field hypothesis is quite realistic, for simplicity, we will use the slow-roll approximation, i.e.,
\begin{equation}\label{kleina}
3H\dot{\phi} + \frac{dV}{d \phi} = 0\;,
\end{equation}
\begin{equation}\label{fria}
{{H}}= \frac{1}{\sqrt{3}} [V(\phi)]^{1/2}\;.
\end{equation}
In addition, we will also neglect the term $\dot{\phi}_{c}^{2}$ in Eq. (7). Thus,
\begin{equation}\label{potentiala}
V(\phi) = V_0 - \sqrt{V_0}\lambda (\phi - \phi_{0}) + \frac{1}{4}\lambda^{2} (\phi - \phi_{0})^{2}\;.
\end{equation}
By combining Eqs. (\ref{kleina}), (\ref{fria}) and (\ref{potentiala}), one finds
\begin{equation}
\phi = \phi_0 + \frac{\sqrt{3}}{3}\lambda (t - t_0)\;.
\end{equation}
From previous equation, it follows that
\begin{equation}
\dot{\phi} = \frac{\sqrt{3}}{3}\lambda\;,
\end{equation}
which completely recovers the assumption given by Eq. (5). 

\begin{figure*}
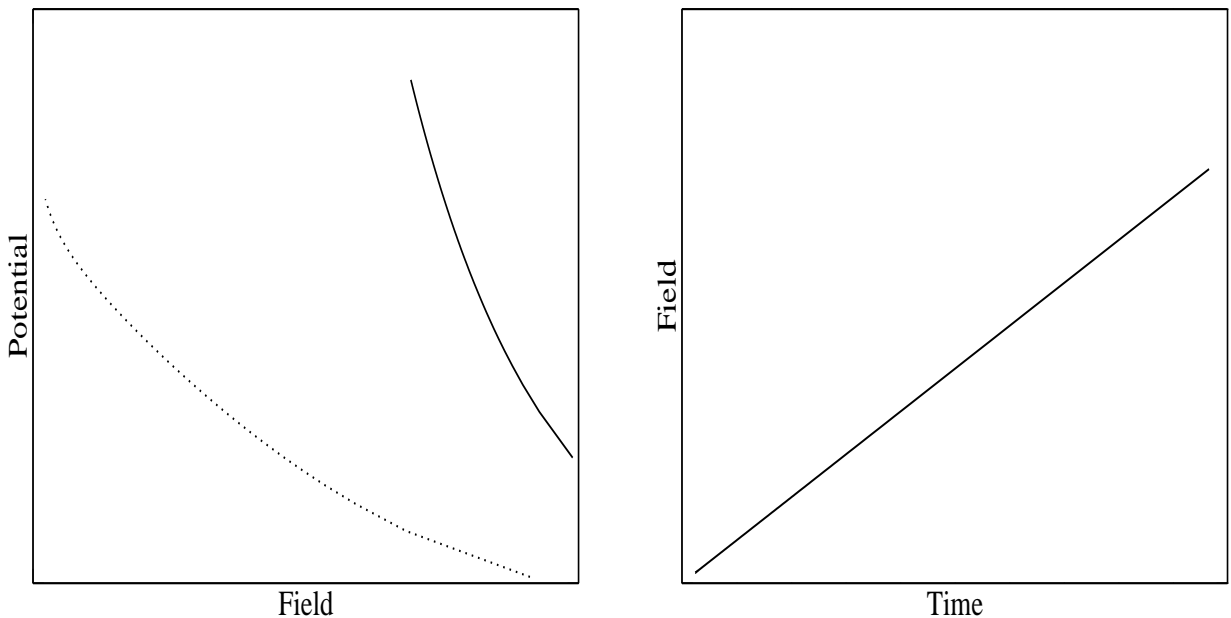

\centerline{\psfig{figure=fig6.eps,width=3.0truein,height=3.2truein}
\hspace{0.9cm}\psfig{figure=fig7.eps,width=3.0truein,height=3.2truein}}
\caption{Left panel: Behaviour of the potential as funtion of the field (without matter - full line) and (with matter - dotted line). Right panel: Evolution of $\phi(t)$ for general case in which the matter is included.}
\label{fig:qzw}
\end{figure*}

\subsection{Terminal scalar field and matter model}

From now on, it will be considered that the cosmological fluid is composed of non-relativistic matter (baryonic plus dark) and a slow rolling scalar field with constant kinetic term. Since that these components are separately conserved, one has that
\begin{equation}\label{evomatter}
\dot{\rho}_{m} + 3\frac{\dot{a}}{a}\rho_{m} = 0 \Longrightarrow \rho_{m} = \rho_{m,0}a^{-3}\;,
\end{equation}
and
\begin{equation}\label{evopfi}
\dot{\rho}_{\phi} + 3\dot{\phi}_{c}^2 \frac{\dot{a}}{a} = 0 \Longrightarrow \rho_{\phi} = \rho_{\phi,0} + 3\dot{\phi}_{c}^2 \ln a^{-1}\;,
\end{equation}
where $\rho_{m,0}$ and $\rho_{\phi,0}$ are, respectively, the current values of the energy densities of the matter and of the scalar field. Now, the expansion rate of the Universe can be written as
\begin{equation}\label{taxat}
H= H_0 {\left[\Omega_{m}a^{-3} + \Omega_{\phi} + 3\Omega_{c} \ln a^{-1}\right]}^{1/2}\;,
\end{equation}
where $\Omega_{m} =\rho_{m,0}/3H_{0}^{2}$ and $\Omega_{\phi} = \rho_{\phi,0}/3H_{0}^{2}$. It is interesting to observe that this model differ of the standard model ($\Lambda$CDM) only by adding of the term $(3\Omega_{c} \ln a^{-1})$. As $\ln a^{-1}$ is very small for any $a \leq 1$, all cosmic evolution from past to present, in this background, is very similar to $\Lambda$CDM (see figure 2 $-$ left panel, where the curves are overlapped). However, as shown in figure 2 $-$ right panel, the future evolution from our model differs completely of the $\Lambda$CDM model. In our case when $z \longrightarrow - 1 \Longrightarrow H/H_0 \longrightarrow 0$ and the Universe collapses.

Now, from relation $[\rho_{\phi} = (1/2){\dot{\phi}}_{c}^{2} + V(\phi)]$, it follows that
\begin{equation}\label{vdea}
V(a) = \tilde{V}_{0} + 3\dot{\phi}_{c}^{2} \ln a^{-1},
\end{equation}
where $\tilde{V}_0 = \rho_{\phi,0} - (1/2){\dot{\phi}}_{c}^{2}$.
On the other hand,
\begin{equation}\label{fiacampo}
\phi = \dot{\phi}_{c}t = \dot{\phi}_{c} \int \frac{1}{H}\frac{da}{a}.
\end{equation}
By combining numerically Eqs. (\ref{taxat}), (\ref{vdea}) and (\ref{fiacampo}) we find the resulting potential $V(\phi)$.

Figure 3 (left panel) shows the evolution of the potential obtained numerically (dotted line) and analytically (full line). We can see that the presence of matter only decreases the slope of the potential. Physically, this amounts to saying that the field rolls more slowly in the presence of material content than in the absence of matter, so that the general potential belongs to the same class of potentials given in Eq. (\ref{potential}). 

For the sake of completeness, we also show in figure 3 (right panel) the field $\phi$ as a function of time. In this case the slope of the curve is constant, so that we conclude that $\dot{\phi} = constant$.

\section{Observational Analysis}

Now, we will discuss the observational bounds on the free parameters of the model previously discussed. To this end, we perform a joint analysis involving different observational sets of data, as described following.

\begin{figure*}
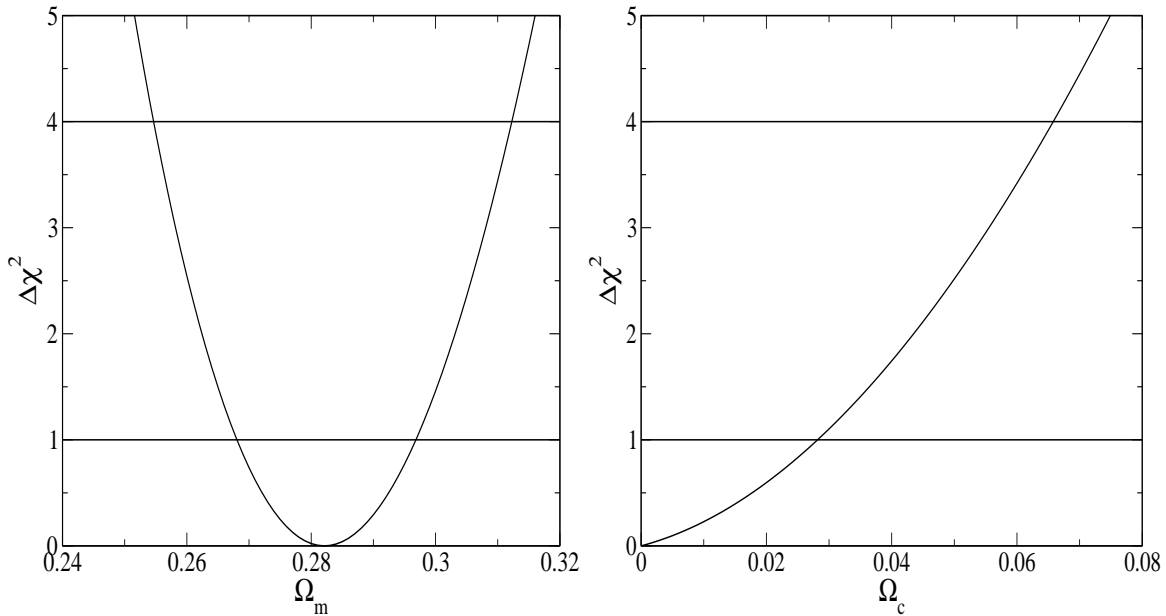

\centerline{\psfig{figure=fig4.eps,width=3.0truein,height=3.2truein}
\psfig{figure=fig5.eps,width=3.0truein,height=3.2truein}}
\caption{The results of our statistical analysis. The variance $\Delta \chi^2$ as a function of the parameters $\Omega_{m}$ (left panel) and $\Omega_{c}$ (right panel) at 2$\sigma$ confidence level.}
\label{fig:qzw}
\end{figure*}

We use one of the latest supernovae catalog compiled by (Suzuki et al. 2012) which includes 580 data points after selection cuts. The best fit of the parameters $s$ = ($\Omega_{m}$, $\Omega_{c}$) is found by using a $\chi^2$ statistics. Thus,
\begin{equation}\label{chisquare}
\chi^2_{\rm{SN}} = \sum_{i=1}^N\frac{{[\mu_{p}^{i}(z|s)(z_i) - \mu_{o}^{i}(z|s)}]^{2}} {\sigma_i^2}\;,
\end{equation}
where $\mu_{p}^{i}(z|s) = 5\log d_L + 25$ is the predicted distance modulus for a supernova at $z$, $d_L$ is the luminosity distance, $\mu_{o}^{i}(z|s)$ is the extinction corrected distance modulus for a given SNe Ia at $z_i$ and $\sigma_i$ is the uncertainty in the individual distance moduli.

We also use measurements derived from the product of the CMB acoustic scale and from the ratio of the sound horizon scale at the drag epoch to the BAO dilation scale, i.e.,
\begin{equation}
f_{z_{BAO}} \equiv \frac{d_A (z_*)}{D_V(z_{\rm{BAO}})} \frac{r_s(z_d)}{r_s(z_*)}\;,
\end{equation}
where $d_A (z_*)$ is the comoving angular-diameter distance to recombination ($z_* = 1090$), $D_V$ is the dilation scale, $r_s$ is the comoving sound horizon at photon decoupling and $z_d \simeq 1020$ is the redshift of the drag epoch (at which the acoustic oscillations are frozen in). For $z_{\rm{BAO}} =$ 0.2, 0.35 and 0.6., one 
finds $f_{0.2} = 18.32 \pm 0.59$, $f_{0.35} = 10.55 \pm 0.35$ and $ f_{0.6} = 6.65 \pm 0.32$ (Sollerman et al. 2009; Blake et al. 2011) [see also (Percival 2010)].

In our analysis, we minimize the function $\chi^2_{\rm{T}} = \chi^2_{\rm{SN}} + \chi^2_{\rm{BAO/CMB}}$, where $\chi^2_{\rm{BAO/CMB}}$ correspond to the BAO/CMB $\chi^2$ function. 

Figure 4 shows the variance $\Delta \chi^{2} = \chi^{2} - \chi^{2}_{min}$ as a function of the parameters $\Omega_{m}$ (left panel) and $\Omega_{c}$ (right panel) at 2$\sigma$ confidence level.
From this analysis, we find $\Omega_{m} = 0.282_{-0.03}^{+0.03}$ and $\Omega_{c}= 0.0_{-0.00}^{+0.07}$. Therefore, in excellent agreement with observations.

\section{Conclusions}

We have developed a cosmological scenario from an analogy between the dynamics of a scalar field with constant kinetic term and the vertical movement of a skydiver after reaching terminal velocity. We have found an analytical expression for the scalar field potential that is given by a quadratic function of the field. We have shown that the past evolution of the Universe to present predicted by our scenario is similar to the standard model, however, the future cosmic evolution is completely different. While ($\Lambda$CDM) model predicts that Universe evolves to de Sitter phase, our model predicts that the future of the Universe will be followed by a contraction phase. The collapse time depends only on $\Omega_c$ and $H_0$ parameters and is estimated in 190 Gyears from today. Finally, we have performed a joint statistical analysis with different observational data and have found stronger constrains on free parameters of our model.

\begin{acknowledgments}

This work was supported by CNPq (Brazilian Research Agency) No 453848/2014-1 and UFERSA No 8P1618-22. The author thanks to J. M. F. Maia for valuable discussions.

\end{acknowledgments}

\label{lastpage}
\end{document}